\documentclass[twocolumn,showpacs,preprintnumbers,amsmath,amssymb,prl]{revtex4}
\usepackage{amssymb}
\usepackage{graphicx}
\usepackage{amsmath}
\usepackage{epsfig}

\usepackage[draft]{hyperref}

\begin{document}

%\date{\today : version 1.1}
\title{\textbf{Coherent Amplification of Optical Pulses in Metamaterials}}

\author{Ildar R. Gabitov }
\address{Department of Mathematics, University of Arizona, 617 North Santa Rita Avenue, Tucson, AZ 85721, USA\\
Email:  gabitov@math.arizona.edu}
\author{Bridget Kennedy}
\address{Department of Mathematics, University of Arizona, 617 North Santa Rita Avenue, Tucson, AZ 85721, USA \\
Email:  bkennedy@math.arizona.edu }
\author{Andrei I. Maimistov }
\address{Department of Solid State Physics and Nanosystems, Moscow Engineering Physics Institute, Kashirskoe sh. 31, Moscow 115409, Russia \\
Email: maimistov@pico.mephi.ru}

%\pacs{42.25.Bs, 42.65.Pc, 42.81.Dp,}

\begin{abstract}
In this paper we theoretically study propagation of steady state
ultrashort pulse in dissipative medium. We considered two cases (i)
medium  consists of lossy  metallic nanostructures embedded into a
gain material and (ii) the gain material is embedded directly into
the nanostructures. We found the shape and velocity of an optical
pulse coupled with the polarization wave.
\end{abstract}

\maketitle

\section{Introduction}
\noindent For over a decade extensive research has been undertaken
on negative refractive indexed metamaterials~\cite{
Shelby01,Shalaev05,Valentine08} both in the microwave and optical
regimes.  The research in optical metamaterials has been stimulated
by the possibility of light manipulation methods and potential
applications.   Most of the experimental work with metamaterials
utilizes the plasmonic resonance in metallic structures embedded
into a host dielectric.   The  disadvantage of this type of
metamaterial arises from the high loss value that results from an
induced plasmonic resonance in the optical range. Currently, the
reduction  or compensation of these losses has been the the focus of
intensive research.  Conceptually, the compensation of losses can be
achieved in two ways.  First, by adding gain into a host material
containing the metallic nanostructures.  Second, by adding a gain
material into the nanostructures themselves.  In this paper, we
consider the optical pulse dynamics in a negative index metamaterial
using both types of loss compensation techniques.  To understand the
general features of the pulse dynamics in such a system we consider
a simplified mathematical model.

\section{Basic Equations}
The propagation of an electromagnetic wave in matter is described by the system of Maxwell's equations:
\begin{eqnarray}
\nabla \times \vec{E} +\frac{1}{c} \frac{\partial \vec{B}}{\partial t} =0,~~~~ \nabla \times \vec{H} -\frac{1}{c}\frac{\partial \vec{D}}{\partial t}=0 .
\label{Vector:Maxwell}
\end{eqnarray}
By choosing a system of coordinates in such a way as to assume that a wave propagates in the $z$-direction, the magnetic field is oriented in $y$-direction,  $\vec{H} =\left (0, H(z, t), 0\right)$, and the electric field is oriented in the $x$-direction,  $\vec{E} =\left(E(z,t), 0, 0\right)$, we transform the system of vector equations~(\ref{Vector:Maxwell}) into a system of scalar equations:
\begin{eqnarray}
\frac{\partial E}{\partial z}  +\frac{1}{c} \frac{\partial B}{\partial t} =0,~~~~\frac{\partial H}{\partial z}  +\frac{1}{c}\frac{\partial D}{\partial t}=0 .
\label{scalar:Maxwell}
\end{eqnarray}
We start our analysis of the system from the initial case of an electromagnetic wave as it propagates through a nanostructured host dielectric without a gain material.  We also assume that all resonances of the host material are far from the resonances of the metallic nanostructures.  In this case,  $D$ and  $B$ can be represented in the standard form through the medium polarization, $P$, and the magnetization, $M$:
\begin{equation}
B=H+4 \pi M,~~~D=E+4 \pi P.
\label{B:and:D:general}
\end{equation}

To close the system of equations we need to add material equations that describe the interaction of $P$ and $M$ with an electromagnetic wave.    The total medium response can be represented as sum of the responses from the nanostructures and the gain material: $P=P_{ns}+P_g$ and $M=M_{ns}$.  To describe such an interaction, we consider the simplified model represented by distributed linear electric and magnetic resonators. The plasmonic oscillations result in an oscillation of the polarization described by following equation~\cite{Gabitov06}:
\begin{eqnarray}
\frac{\partial^{2} P_{ns}}{\partial t^2} +\gamma\frac{\partial P_{ns}}{\partial t}+\omega^2 _D P_{ns}  =\frac{\omega^2 _p}{4\pi} E.
\label{polarization_eqn}
\end{eqnarray}
Here  $\omega_p$ is the effective plasma frequency,  $\omega_D$ is the frequency of the dimensional quantization that results from the geometry of the resonator,  and $\gamma$ accounts for losses. In the popular fishnet structures the dielectric permittivity is negative in operating frequency range. In this case we can choose   $\omega_D =0$.

To describe the magnetic response we assume that a magnetic resonator in a leading order  operates as an $LC$-circuit~\cite{Tamm79}.   This allows us to write a partial differential equation that relates the magnetization and the magnetic field as follows~\cite{Markos062}:
\begin{eqnarray}
\frac{\partial^{2} M_{ns}}{\partial t^2}+\alpha \frac{\partial M_{ns}}{\partial t}+ \omega^2_T M_{ns} = -b \frac{\partial^{2} H}{\partial t^2},
\label{magnetization_eqn}
\end{eqnarray}
where $L$, $C$  are effective inductance and capacity,  $\omega^2_T$ is a Tompson frequency, $\alpha$ is a dissipative constant, and $b
$ describes coupling of nanoresonator with external magnetic flux~\cite{Smith:PhysRevLett.84.4184}.

The system~(\ref{scalar:Maxwell}), (\ref{B:and:D:general}), (\ref{polarization_eqn}),    (\ref{magnetization_eqn}) is a generalization of the well known Maxwell-Lorentz system of equations, which describes the resonant interaction of an electromagnetic field with a system of linear oscillators that responds only to the electric component of incident light.   This system has played a fundamental role in understanding the interaction of light with resonant atoms~\cite{Lorentz,AE87}.

The host material is characterized by the dielectric permittivity, $\varepsilon_{0}(\omega)$, and the magnetic permeability, $\mu_{0}(\omega)$. For most materials the permeability in the optical range is $1$, that is, $\mu_0(\omega)=1$.  Now, by taking the Fourier transform of the equations in~(\ref{scalar:Maxwell}) and substituting in the expressions for $\hat{B}$ and $\hat{D}$
we  obtain the following equations for Fourier components $\hat{E}$ and $\hat{H}$ in terms of $\hat{P}$ and $\hat{M}$:
\begin{eqnarray}
k^2 \hat{E} -\frac{\omega^2 \varepsilon_0(\omega) }{c^2}\hat{E}&=& \frac{4\pi  \omega^2}{c^2}\hat{P}+  \frac{4\pi  \omega}{c} k\hat{M} \notag \\
k^2 \hat{H} -\frac{\omega^2 \varepsilon_0(\omega)}{c^2} \hat{H}&=&\frac{4\pi  \omega^2 \varepsilon_0(\omega)}{c^2}\hat{M}+  \frac{4\pi \omega}{c} k\hat{P}.
\label{Fourier:Maxwell}
\end{eqnarray}

In order to describe the response from the nanostructures we introduce both an effective dielectric and a magnetic susceptibility.  We consider here the case of a "diluted" metamaterial when near neighbor interactions of the nanostructures are much weaker than their interaction via the electromagnetic field.  The nanostructure contribution to the effective permittivity and permeability can be obtained from equations~(\ref{polarization_eqn}) and~(\ref{magnetization_eqn}) by applying the Fourier transform, that is,
\begin{eqnarray}
\hat{P}_{ns} =\frac{\omega^2_p/4\pi}{(\omega^2_D -\omega^2)-i\omega \gamma}\hat{E}=\chi_e(\omega)\hat{E}\label{polarization},\\
\hat{M}_{ns} =\frac{b \omega^2}{(\omega^2_T -\omega^2-i\omega \alpha)}\hat{H}=\chi_m(\omega)\hat{H}\label{magnetization}.
\end{eqnarray}

The effective permittivity, $\tilde{\varepsilon}$, and the permeability, $\tilde{\mu}$,  can therefore be expressed
\begin{eqnarray}
\tilde{\varepsilon}(\omega)&=&\varepsilon_{0} +4 \pi \chi_{e}(\omega),\nonumber\\
\tilde{\mu}(\omega)&=&1 +4 \pi \chi_{m}(\omega)
\label{epsilon:mu}
\end{eqnarray}
Using~(\ref{epsilon:mu}), the system of equations~(\ref{Fourier:Maxwell}) can be represented as
\begin{eqnarray}
\left[ k^2 -\left(\frac{\omega }{c}\right)^{2} \tilde{\varepsilon}(\omega)\tilde{\mu}(\omega)\right]\hat{E}(k,\omega)=~~~~~~~  \nonumber\\
~~~~~4\pi \left(\frac{\omega }{c}\right)^{2} \tilde{\mu}(\omega) \hat{P}_{g}(k,\omega)~~~ \label{Fourier:new:E}\\
\left[ k^2 -\left(\frac{\omega }{c}\right)^{2} \tilde{\varepsilon}(\omega)\tilde{\mu}(\omega)\right]\hat{H}(k,\omega)=~~~~~~~ \nonumber\\
 ~~~~~4\pi  \frac{\omega}{c}k \hat{P}_{g}(k, \omega).~~~  \label{Fourier:new:H}
\end{eqnarray}

Ultrashort optical pulses  are well described by a slowly varying envelope approximation.  This approximation allows one to simplify the system  ~(\ref{scalar:Maxwell}), (\ref{B:and:D:general}), (\ref{polarization_eqn}),   (\ref{magnetization_eqn}) by representing $E$, $H$, $P$ and $M$ in the following form:
\begin{eqnarray}
E(z, t) &=& \mathcal{E}(z, t) \exp{(i\omega_0 t  -i k_0 z)} + c.c.,\nonumber\\
P(z,t) &=& \mathcal{P}(z, t) \exp{(i\omega_0 t -i k_0 z)} + c.c.,\nonumber\\
H(z, t) &=& \mathcal{H}(z, t) \exp{(i\omega_0 t -i k_0 z)} + c.c.,\label{factorization}\\
M(z, t) &=&\mathcal{M}(z, t) \exp{(i\omega_0 t -i k_0 z)} + c.c., \notag
\end{eqnarray}
where  $\mathcal{E}(z, t)$, $\mathcal{P}(z, t)$, $\mathcal{H}(z, t)$ and $\mathcal{M}(z, t) $  are slowly-varying functions of $t$ and $z$. It should be noted that the constant $k_0$ stands for the projection of the wave vector on $z$ axis, and that it can be both positive and negative for positive and negative index materials respectively.

The factorization~(\ref{factorization}) means that the spectral width of the signal is much smaller than carrier frequency, therefore the equations for the slowly varying amplitudes can be obtained by a frequency and wave vector shift, $\omega \rightarrow \omega +\omega_0,~~k\rightarrow k_{0}+k$, in Fourier space~\cite{maimistov99}. Thus, the  equations describing the slow dynamics read
\begin{eqnarray}
\left[ \tilde{k}^2 -\left(\frac{\tilde{\omega} }{c}\right)^{2} \tilde{\varepsilon}(\tilde{\omega})\tilde{\mu}(\tilde{\omega})\right]\mathcal{\hat{E}}(k,\omega)=~~~~~~~~~~ \nonumber\\
~~4\pi \left(\frac{\omega }{c}\right)^{2} \tilde{\mu}(\omega) \mathcal{\hat{P}}_{g}(k,\omega)&~& \label{Fourier:slow:E}\\
\left[ \tilde{k}^2 -\left(\frac{\tilde{\omega} }{c}\right)^{2} \tilde{\varepsilon}(\tilde{\omega})\tilde{\mu}(\tilde{\omega})\right]\mathcal{\hat{H}}(k,\omega)=~~~~~~~~~  \nonumber\\
~~ 4\pi  \frac{\tilde{\omega}}{c}\tilde{k} \mathcal{\hat{P}}_{g}(k, \omega),    \label{Fourier:slow:H}\\
\tilde{\omega}=\omega+\omega_{0},~~~\tilde{k}=k+k_{0}.~~~~~~~~~~~~~~~~~~~ \notag
\end{eqnarray}
Here the parameter $k_0$ can be both positive and negative, since it represents the projection of the wave vector of the carry wave on the direction of the wave propagation.  In the case of a negative refractive index the sign of  $k_0$ is negative.

Since the width of wave packet is much less than carrier frequency, the inequalities $\omega \ll \omega_0$ and $k\ll |k_0|$ still hold.  By expanding the expression in the square brackets in equations ~(\ref{Fourier:slow:E}) we obtain
\begin{eqnarray}
[\ldots] &\approx & k^2_0 +2k_0 k+k^2 - \frac{(\omega^2_0 +2 \omega_0 \omega +\omega^2)}{c^2}[ n^2_0 (\omega_0)  \nonumber \\
&+& \left[\frac{\partial n^2}{\partial \omega}\right]_{\omega_0} \omega + \frac{1}{2} \left[\frac{\partial^2 n^2}{\partial \omega^2}\right]_{\omega_0} \omega^2],
\end{eqnarray}
where $n^{2}=\tilde{\varepsilon}(\omega)\tilde{\mu} (\omega)$ is a square of the effective complex index of refraction.

The effective index of refraction has both a real and an imaginary part.  The imaginary part is responsible for the phase dynamics of the field, while the imaginary part is responsible for field decay. We consider the propagation constant $k$, which is defined as follows
\begin{equation}
k(\omega) = (\omega /c)\mathrm{Re}\sqrt{\varepsilon (\omega)\mu (\omega)}.
\label{dispersion:law}
\end{equation}
Introducing  the auxiliary  function $\Gamma(\omega)$
\[
\Gamma(\omega)=(\omega /c)\mathrm{Im} \sqrt{\varepsilon (\omega)\mu (\omega)}
\]
and  applying the inverse Fourier transform, we obtain equations in spatio-temporal variables:
\begin{eqnarray}
i\left(\frac{\partial}{\partial z} +\frac{1}{v_g}\frac{\partial}{\partial t} -\frac{i}{2} \left[\frac{\partial^2 k}{\partial \omega^2}\right]_{\omega_0} \frac{\partial^2}{\partial t^2}+ \Gamma_{0}  \right)\mathcal{E}(z,t) \nonumber \\
= \frac{2 \pi \omega_{0} \mu(\omega_0)}{c ~\mathrm{Re~}\sqrt{\mu (\omega_0) \varepsilon(\omega_0)}}\mathcal{P}_{g},~~~~~~~ \label{svepa:E}\\
i\left(\frac{\partial}{\partial z} +\frac{1}{v_g}\frac{\partial}{\partial t}-\frac{i}{2} \left[\frac{\partial^2 k}{\partial \omega^2}\right]_{\omega_0} \frac{\partial^2}{\partial t^2}+ \Gamma_{0} \right) \mathcal{H}(z,t) \nonumber  \\
= 2 \pi \frac{\omega_0}{c}\mathcal{P}_{g}.~~~~~~
\label{svepa:H1}
\end{eqnarray}
Here the group velocity $v_g$ is defined as
$  v_{g}^{-1}=\left[{\partial k}/{\partial \omega}\right]_{\omega_0} $
and $\Gamma_{0}=\Gamma (\omega_0)$.

Note that in the above when we divide both sides by $k_0$ the expansion in the Fourier domain is of second order, where the second order contribution comes from $\omega^2$ term.  In our further analysis we will ignore this dispersive term by dropping this second order term.  Thus, we have the following slowly-varying governing equations for the system:
\begin{eqnarray}
i\left(\frac{\partial}{\partial z} +\frac{1}{v_g}\frac{\partial}{\partial t}+ \Gamma_{0}  \right)\mathcal{E}(z,t) =\nonumber \\
\frac{2 \pi \omega_{0} \mu(\omega_0)}{c ~\mathrm{Re~}\sqrt{\mu (\omega_0) \varepsilon(\omega_0)}}\mathcal{P}_{g}, \nonumber \\
i\left(\frac{\partial}{\partial z} +\frac{1}{v_g}\frac{\partial}{\partial t}+ \Gamma_{0}  \right) \mathcal{H}(z,t)=
2 \pi \frac{\omega_0}{c}\mathcal{P}_{g}.
\label{svepa:H}
\end{eqnarray}

It follows from equations~(\ref{svepa:H}) that the magnetic field is proportional to the electric field.  Therefore, it is sufficient to analyze only the equation for electric field.  Also, note that if the carrier frequency is in the frequency range that determines a positive real value for the phase, then $k_0>0$ and the imaginary coefficients will all have a sign opposite that which that they are for the case when $k_0<0$.

\section{Amplifying host media: slowly-varying approximation}
As noted in the previous section, the metallic  nanoresonators are lossy.  To compensate for these losses we  propose that active atoms be embedded in the dielectric host medium or into nanostructures. It should be noted that here we consider coherent amplification, where the coherent state of the atomic ensemble  is preserved during transition from an excited to a ground state.  The atomic inclusions that we consider here will be two-level, that is, they have an excited and a ground state.  In the case when amplifying atoms are included to host dielectric, the polarization, $\mathcal{P}_g$, is represented as polarizability, $\mathcal{P}_{at}$, of a two-level single multiplied by atomic density, $n_{at}$: $\mathcal{P}_g = n_{at} \mathcal{P}_{at}$. The Bloch equations, which describe the interaction of polarizability and population difference, $\mathcal{N}$, of two-level atoms with the electric field is given in the slowly-varying envelope approximation as:
\begin{eqnarray}
\frac{\partial N}{\partial t} &=& \frac{2i}{\hbar}(\tilde{\mathcal{E} }^*\tilde{\mathcal{P} }_{at} - \tilde{\mathcal{E} }\tilde{\mathcal{P} }^*_{at})\\
\frac{\partial \tilde{\mathcal{P} }_{at}}{\partial t} &=& \frac{i d^2} {\hbar} N\tilde{\mathcal{E} }-i(\omega_{12} -\omega_0) \tilde{\mathcal{P} }_{at},
\end{eqnarray}
where $\omega_0$ is the carrier frequency, $\omega_{12}$ is the frequency corresponding to the transition between the excited and ground state, $d$ is the dipole moment.

We consider the case when the spectral pulse width, $\Delta \omega$, is within the negative index frequency, $\Delta \omega_{NI}$, domain and that the pulse width, $\tau_0 \sim (\Delta \omega)^{-1}$, is smaller than the characteristic time of the atomic polarization relaxation, $T_2$.  In an experimental situation, $\Delta \lambda_{NI} \sim 50nm$~\cite{Chettiar07,Chettiar08}, the value of $T_2$ varies in a broad range.  For $Er$ ions in a glass matrix at room temperature, $T_2 \sim 250fsec$ and $T_2 \sim 3.6\mu sec$ for $LaF_{3}:PR^{3+}$ at $2^{\circ}K$~\cite{DeVoe79}.  We consider the situation when all these requirements are fulfilled and coherent amplification takes place.

To fulfill these requirements and select an optimal value for the carrier frequency we analyze the dispersion relation~(\ref{dispersion:law}). The negative index property is determined by the magnetic response of the metamaterial. The parameter $b$ determines the strength of this response and therefore controls the negative index property.  This parameter depends on the design of the nanostructure.  The case of a double split ring was considered in~\cite{Obrien04,PhysRevLett.95.237401}, and from these papers it follows that the value of $b$  can be chosen from the interval $[0.1- 0.7]$.  In real experimental situations of doubly resonant metamaterials, the values of  $\omega_D$ and $\omega_T$ are close.  In the case of a fishnet structure there is no resonance in the electric response and $\omega_D$ can be chosen to be zero $\omega_D =0$~\cite{Obrien04}.  In our study we chose  $b = 0.35$ and all the frequencies are normalized relative to $\omega_T$.  Thus we chose  $\omega_D =1.1$,  $\omega_T =1$, $\omega_p =4$ and the dissipative constants were chosen as $\gamma  =0.003$ and $\alpha = 0.03$.  This choice of parameters approximately corresponds to the case of silver.  The value of $\gamma$ is taken from~\cite{PhysRevLett.95.237401} and the value of $\alpha$ was extracted from experimental results presented in~\cite{Chettiar07,Chettiar08}.
\begin{figure}
\begin{center}
\includegraphics[scale=0.4]{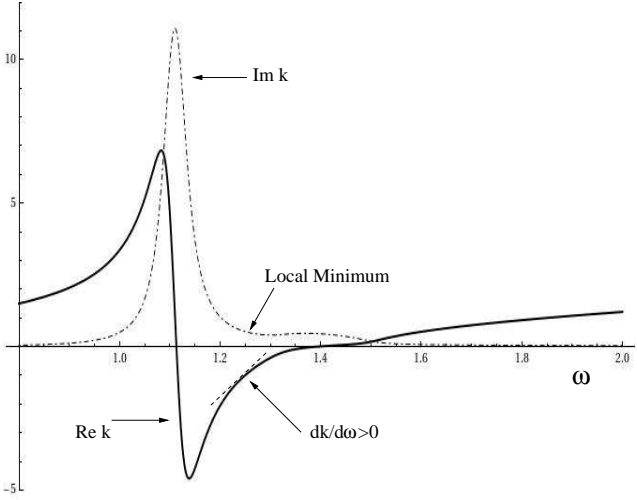}
\end{center}
\caption{The dispersion relation}
\label{dispersion:relation}
\end{figure}
The functions  $k_0(\omega)$ and $\Gamma (\omega)$ are presented in Figure~\ref{dispersion:relation}.  The function $k_0(\omega)$ is negative on the interval $[1.1-1.3]$.  The function  $\Gamma (\omega)$ is positive, which corresponds to the presence of losses, and  has a flat local minima around $\omega \sim 1.2$.  The neighborhood of this flat local minima of $\Gamma (\omega)$ and interval of negative values for $k_0(\omega)$ overlap.  If the  carrier frequency, $\omega_0$, is chosen near this local minima then the effect of dispersion of dissipation can be neglected.   Since $k_{0}(\omega_0)<0$, the phase velocity  is negative, while on the other hand the group velocity, $v_g$, is positive $\left(\partial k_0 / \partial \omega \right)_{\omega_0} >0 $ for this interval of frequencies.  Therefore, the phase and group velocities are oppositely directed for $\omega_0$ selected as described above.   If the resonant frequency of the active atoms is close to such  an $\omega_0$, the material possesses the negative index property and the dispersion of losses can be neglected.

\subsection{Analytic Computation of the Front}
The balance between coherent amplification and dissipation leads to the formation of a  stationary optical pulse. The shape and velocity of such pulse can be found using technique suggested in~\cite{Man:Kam87,Man:Kam90}. In the following dimensionless variables:  $\mathcal{E}=E_0e$, $\mathcal{P}_g=P_0 p$, $N=N_0n$, $\zeta=z/\zeta_0$, $\eta=t/\eta_0$ we can obtain the following governing equations for the system
\begin{eqnarray}
e_\zeta + e_\eta +\gamma e &=&(1+i\delta)p \notag\\
n_\eta& =& -\frac{1}{2}(e^*p +ep^*) \label{Dimensionless:PDE}\\
p_\eta+i\lambda p& =&ne, \notag
\end{eqnarray}
where
\begin{eqnarray*}
 \gamma=\Gamma_0\zeta_0,~ , ~
\delta=\frac{\mathrm{Im}(\mu(\omega_0))}{\mathrm{Re}(\mu(\omega_0))},  ~
\zeta_0=\eta_0 v_g, ~ \\
P_0=N_0, ~E_0=\frac{\hbar}{4\eta_0},  ~
P_0=\frac{c\hbar \mathrm{Re}(\sqrt{\mu(\omega_0)\epsilon(\omega_0)})}{(2\pi\omega_0 \tau^2_0 v_g \mathrm{Re}(\mu(\omega_0)))}.
\end{eqnarray*}

By examining the behavior of the system near the excited state, $n=1$, we will be able to study the linear behavior of the front.  The system of equations near the excited state has the following approximation in the comoving system of coordinates   $\tau=\eta-\zeta/v$, $z=\zeta$:

\begin{eqnarray}
-\beta e_\tau +e_z +\gamma e=(1+i \delta) p \nonumber \\
p_\tau + i\lambda p = e.
\end{eqnarray}
where $\beta = (1-v)/v$.  Taking the Fourier transform of this system gives us the dispersion relation:
\begin{eqnarray}
k(\omega) = -\beta  \omega  -i\gamma +\frac{1+i \delta}{\omega +\lambda}
\end{eqnarray}
Thus, near the excited state an amplitude of  the electric field can be approximated as
\begin{eqnarray}
e(z, \tau) =\int_{-\infty}^{\infty} \hat{e}_0(\omega)\exp{\left(i\omega \tau -i \Theta(\omega)z\right)}d\omega,
\label{e:integral}
\end{eqnarray}
where  $\Theta(\omega) =  -\beta  \omega -i\gamma +(1+i \delta)/(\omega +\lambda)$ is the phase of the integral.
We are interested in the behavior of the traveling wave solution for large distances so as to determine solitary wave solutions.  The method for determining this type of a solution for integrals of this form is the stationary phase technique.
The phase, $\Theta(\omega)$, has a fixed point at:
$-(1+i\delta)/\beta=\omega^2$.  That is,
\begin{eqnarray}
\omega_{\pm} = -\lambda \pm i \sqrt{\frac{1+i\delta}{\beta}}.
\end{eqnarray}
For these values of $\omega$ we see that the phase has the following forms:
\begin{eqnarray}
\Theta(\omega_{\pm}) = \lambda \beta  -i \gamma \mp 2 i \sqrt{1+i\delta}\sqrt{\beta }
\end{eqnarray}
Thus, we see that when the positive square root is chosen the solution will grow exponentially for large $z$, which cannot occur, and thus the contribution from this term must be zero.  For the second stationary point there is an exponential decay, which implies that we will have a strong dependence on $z$.  In our traveling wave solution, the shape of the pulse for large $z$ should be determined by $\tau$, and therefore this strong dependence of $z$ must be eliminated.  Thus, we need the following to hold:
$
2 $Im$[i\sqrt{1+i\delta}]\sqrt{\beta} - \gamma=0,
$
that is,
\begin{eqnarray*}
\beta_{cr}= \frac{\gamma^2}{4\mathrm{Im}[ i\sqrt{1+i\delta}]^2}\\
v_{fr}=\frac{4 \mathrm{Im}[ i\sqrt{1+i\delta}]^2}{4\mathrm{Im}[ i\sqrt{1+i\delta}]^2+\gamma^2} \nonumber
\end{eqnarray*}

Using this expression for $\beta_{cr}$ and substituting in $\omega_+$ into the phase of the integral in expression~(\ref{e:integral}) we see that near the excited state the electric field can be approximated as follows:
\begin{eqnarray}
e(z, \tau) \approx \hat{e}_0(\omega_+) \exp{(i \omega_+ \tau- i \mathrm{Re}[\Theta(\omega_+)] z)} I(z),
\label{formoffront}
\end{eqnarray}
where $I(z)$ is the integral of higher order terms from the expansion of the phase around that stationary point $\omega_+$.  Note that
$\mathrm{Re}[\Theta(\omega_+)]= \lambda \beta_{cr}+ 2 \mathrm{Re}[i\sqrt{1+i\delta}\sqrt{\beta_{cr}}]$.  Let us use the following notation to simplify this expression:  $\sqrt{1+i\delta} = \delta'+i\delta''$, which allows us to simplify this term:
\begin{eqnarray}
\mathrm{Re}[\Theta(\omega_+)]= \frac{\lambda \gamma^2 - 4\gamma \delta'' \delta'}{4(\delta')^2}.
\end{eqnarray}
Also note that if $\delta''=0$, i.e. if the coefficient in front of the polarization in the first equation in the system of governing equations is not complex, and if we ignore detuning, $\lambda=0$, then the phase will be zero.
Also, note that the form of the front depends on $\hat{e}_0(\omega_+)$, which is determined by the initial condition for the electric field.  In addition, we see that the shape of the pulse is determined by the variable $\tau$, and the imaginary part of $\omega_+$, that is the shape of the pulse is proportional to $\exp(-$Im$[\omega_+]\tau) =\exp(-(2(\delta')^2/\gamma) \tau)$.

We also see that because $\omega_+$ is complex this solution will oscillate in $\tau$ with period determined by Re$[\Theta(\omega_+)]$.  Thus, as we have noted before, when $\delta=0$ the phase is zero, and no oscillations exist.  When numerical computations are performed this can be observed.

We have obtained two very important pieces of information from the linear analysis.  Firstly, we have determined the velocity, $v_{fr}$, of the front, and for this reason we are now able to solve the ordinary differential equation determined by moving into a co-moving frame of coordinates.  Secondly, we now know that the front oscillates as it propagates, so that before we move into the co-moving frame we must express the electric field and polarization envelopes as:  $e(\tau,z)=a(\tau, z) \exp(i K z)$ and $p(\tau,z)=r(\tau,z) \exp(iKz)$.  Plugging these into Eqs.~(\ref{Dimensionless:PDE})
gives us the following system in the co-moving frame $t=\tau-z/v$

\begin{eqnarray}
(-\beta_{cr})a_t +(iK+\gamma )a =(1+i\delta) r\\
n_t = -\frac{1}{2}(ra^* + r^*a)\\
r_t +i \lambda r = na
\end{eqnarray}
Near $n=1$ we can observe that:
\begin{eqnarray}
(-\beta_{cr})a_t +(iK +\gamma )a =(1+i\delta) r\\
r_t +i \lambda r = a.
\end{eqnarray}
With the velocity determined by our previous analysis and recall that $K=-(\lambda \gamma^2 - 4\gamma \delta'' \delta')/4(\delta')^2$ the characteristic equation for the above system is degenerate.  To see this, let us look at the simpler case when $\lambda=0$, that is, when the carrier frequency is equal to the transition frequency of the atomic inclusions.  For this case we have the following ODE
\begin{eqnarray}
-\frac{\gamma^2}{4} a_{tt} +(i\gamma\delta'' \delta' +\gamma (\delta')^2 +\gamma(\delta')^2) a_t - \nonumber \\
(\delta')^2 ((\delta')^2 +2i\delta' \delta'' -(\delta'')^2))a=0,
\end{eqnarray}
which is degenerate with solutions of the form:
\begin{eqnarray}
a(\tau) = c_1 \exp(k_1 \tau) + c_2 \tau \exp(k_1\tau), \\
k_1=\frac{2((\delta')^2 +i\delta' \delta'')}{\gamma}
\end{eqnarray}
Note that to solve the nonlinear ODE we need both $a(0)$ and $r(0)$.  If we can determine $a'(0)$ from the initial condition given for $a$, then we know $r(0)$.  Unfortunately, the above linear analysis does not allow us to do this because the characteristic equation is degenerate and our solution near the excited state does not allow us to simply express $a'(0)$ in terms of the $a(0)$.
However, if we look at the linearization around $n=-1$, the ground state, we see that:
\begin{eqnarray}
(-\beta_{cr})a' +(iK+\gamma )a = r\\
r' +i \lambda r =-a,
\end{eqnarray}
and using the velocity computed in the previous analysis our characteristic equation tells us that $a(\tau)$ can be expressed as:
\begin{eqnarray}
a(\tau) = C_1 \exp(\kappa_+ \tau) + C_2 \exp(\kappa_- \tau),
\end{eqnarray}
where
\begin{eqnarray}
\kappa_{\pm} = \frac{2(1\pm \sqrt{2})(\delta')^2}{\gamma} +\frac{2i\delta'' \delta'(1\pm \sqrt{2})}{\gamma}
\end{eqnarray}

\begin{figure}
\begin{center}
\includegraphics[width=2.5in]{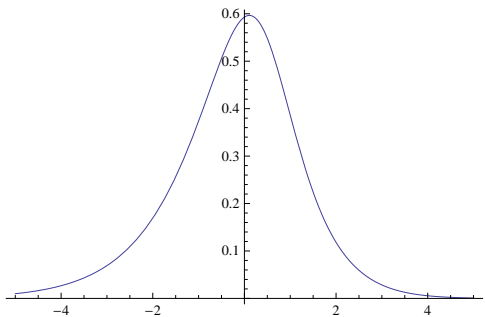}
\end{center}
\caption[Electric field pulse for nonlinear ODE]{This figure represents the numerical integration of the nonlinear ordinary differential equation.  This corresponds to the electric field pulse generated by the transition of our initially excited system.}
\label{electricfield:ODE}
\end{figure}

\begin{figure}
\begin{center}
\includegraphics[width=2.5in]{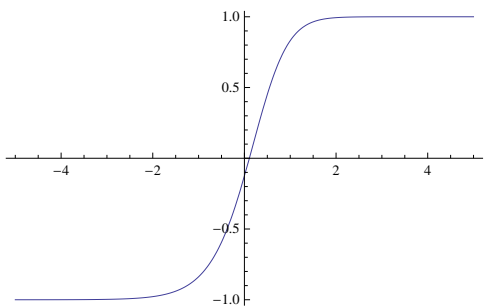}
\end{center}
\caption[Atomic population, $n$, for generation of pulse that corresponds to transition]{This figure represents the relative atomic population as computed by numerical integration of the nonlinear ordinary differential equation.  Note that as the electric field pulse is generated the system goes from the excited to the ground state.}
\label{population:ODE}
\end{figure}

  Since we are analyzing the behavior of the front near the ground state we know that the shape of the pulse must decay exponentially here, and therefore $C_1=0$.  Thus,
\begin{eqnarray}
a(\tau) = a(0)\exp(\kappa_-\tau),~~~~ a'(\tau) =  \kappa_- a(0)  \exp(\kappa_- \tau)
\end{eqnarray}
Thus,  given a choice for $a(0)$ we see that:
\begin{eqnarray}
r(0)= ((-\beta_{cr}) \kappa_- +(iK +\gamma ))a(0)
\end{eqnarray}
and we are able to numerically solve the nonlinear ordinary differential equation.  These solutions can be seen in Figs.~(\ref{electricfield:ODE})-(\ref{population:ODE})

To match the asymptotic behavior let us write the form of the solution near the excited state in the form:
\begin{eqnarray}
a(\tau, z)=e_0(\tau -\tau_{un}) \exp{(k_-(\tau - \tau_{un}))}\exp{iK z}
\end{eqnarray}
This asymptotic form must match the condition obtained using the stationary phase technique.       Since the position of the solution described by the stationary phase technique is described through z while the position of the front solution given by the ODE solution  is determined through the shift variable $\tau_0$, by comparing the asymptotic of these two equations we will be able to express $\tau_0$ as a function of z.

  Thus, if we add a shift term, $\tau_0$ to the ODE solution we can compare asymptotic:
\begin{eqnarray}
\hat{e}_0(\omega_{+}) \exp\left( k_-\tau +iK z\right)I(z)=~~~~~~~~~~ \nonumber \\
e_0(\tau -\tau_0 -\tau_{un}) \exp{(k_-(\tau -\tau_0 - \tau_{un}))}\exp{(iK z)}
\end{eqnarray}
Thus,
\begin{eqnarray}
\hat{e}_0(\omega_{+}) \exp(  k_- \tau )I(z)=~~~~~~~~~~ \nonumber \\
e_0(\tau -\tau_0-\tau_{un}) \exp{(k_-(\tau -\tau_0 - \tau_{un}))}.
\end{eqnarray}
The position of the front is given by $\tau_0$ while the width of the pulse is given by $\tau-\tau_{un}$.  Since the transition from the stable to unstable states takes place on the order $O(1)$ this implies that $\tau-\tau_{un}$ is of order $O(1)$.  On the other hand, $\tau_0$ the position of the pulse is dependent on the size of the incident pulse, $\hat{e}_0$, that is $\tau_0 \sim O(1/\hat{e}_0)$.  Since the initialization is small this means that $\tau_0$ is large.  Thus, $\tau_0>>\tau-\tau_{un}$ and
\begin{eqnarray}
\hat{e}_0(\omega_{+}) I(z)= e_0\tau_0 \exp{( k_- \tau_0) }.
\end{eqnarray}
From this we see that
\begin{eqnarray}
\tau_0 =\frac{1}{k_-}\left(\ln(\hat{e}_0 (\omega_+) I(z) ) -\ln(  \tau_0 e_0 )\right),
\end{eqnarray}
and $\tau_0$, the shift variable, can be determined iteratively.

\subsection{Numerical Computation of Nonlinear PDE}
\begin{figure}
\begin{center}
\includegraphics[width=3.2in]{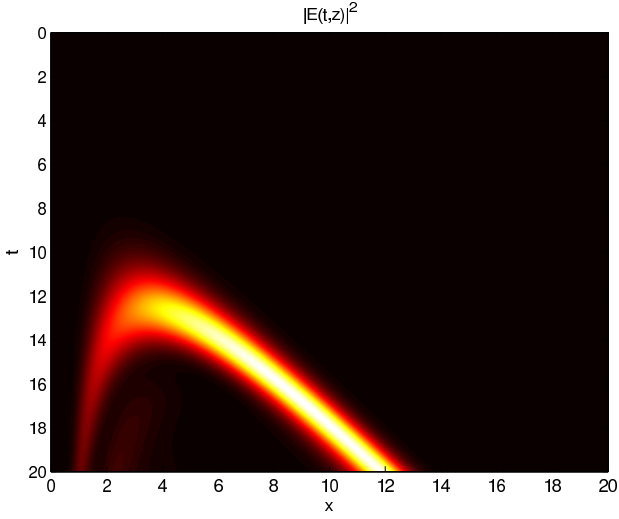}
\end{center}
\caption{This figure showed a plot of the intensity of the pulse.}
\label{Intensity:Numerical}
\end{figure}

Let us return to the dimensionless system of partial differential equations~(\ref{Dimensionless:PDE}).  To solve this system numerically we use a Strang splitting method, where we have a step that solves the ordinary differential equations, and we have a separate step that solves the partial differential dimensionless Maxwell's equation.  The ordinary differential equation step is solved using midpoint method, and the partial differential equation step is solved using the MacCormack method.  At the boundary we consider a small Gaussian pulse $e(t,0)=0.01\exp{(-t^2)}$ and we also assume that the system is initially in the excited state, $n(0,z)=1$.  The result of the numerical computations can be seen in Fig.~(\ref{Intensity:Numerical}).

\begin{figure}
\begin{center}
\includegraphics[width=3.2in]{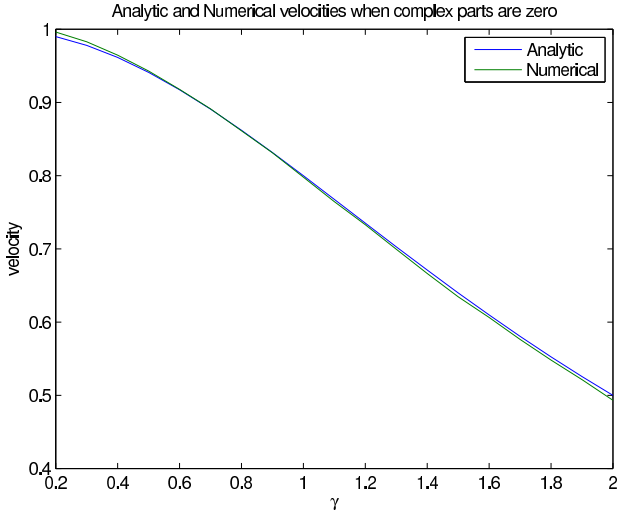}
\end{center}
\caption{This figure shows the result of comparing the velocity of the front generated by the numerical computation of the nonlinear PDE and the velocity of the front that was obtained analytically.}
\label{Analytic:Numerical:complexzero}
\end{figure}

As was noted in the section on linear analysis and in the previous section, the pulse should oscillate as in moves through the medium.  In the numerical computations this does occur, and the period of these oscillations if given by the phase of the approximation of the front given in the expression~(\ref{formoffront}).  A comparison of the velocity computed analytically and the velocity of the pulse generated by the system numerically can be seen in Fig.~(\ref{Analytic:Numerical:complexzero}).

\section{The case of amplifying nanostructures}
Another approach toward loss compensation in materials with embedded nanostructures is to add the amplification capability directly into the nanostructures. There are two main motivations in favor of this approach.  First,  the small size of the nanostructures will induce highly intense local electric fields that result from the interaction of the nanostructures with an external electromagnetic field.  This strong local field will enable a strong response from the gain material if it is inserted directly into a nanostructure.  Secondly, since the nanostructures are the source of loss,  the losses in the system occur on the scale of a nanostructure.  In the case that was considered in the previous subsection, the loss was compensated by a gain material distributed over the entire host medium.  Thus, a strong loss occurred on the scale of a nanostructure, while the compensatation acted on a mean-field scale.  The scales of loss and loss compensation are not in agreement when performed in this way.  A more appropriate way to compensate for loss would not only balance the gain and loss of energy, but would also match the scales of these two procedures.   We propose that this can be achieved by embedding a gain material directly into the nanostructures, and in this subsection we focus on modeling such a method of loss compensation.

In our analysis we consider the simplest case where  nanostructure in a leading order can be viewed as an $LC$ circuit.  We consider  nanostructures that produce spatially homogeneous effective local fields.  This simplification allows us to avoid a consideration of the spatial dynamics for a local field, and to take into account only temporal dynamics.   The external electric field generates a current in the $LC$ circuit.  The current in turn establishes a local electric field  across a "capacitor".  Thus, if we place active atoms or quantum dots in the part of the structure that is responsible for the electric capacity, we will be able to utilize the strong local electric field and directly compensate for loss.    We first derive an equation that describes the oscillatory processes in an $LC$ circuit that contains active atoms in the capacitor.    In the ideal case without losses, the potential difference at the capacitor can be expressed in the form:
\begin{eqnarray}
U_{C}=-U_{L} + \mathcal{E}_{in}.
\end{eqnarray}

Here $U_{L}$ is potential difference on the inductance and $ \mathcal{E}_{in}$ is the electromotive force due to magnetic flux variations through the area of the $LC$ circuit.  We can express $U_{L}$  via the electric field, $E_{c}$, and the  polarization of the medium, $P_{a}$, inside the capacitor:
\begin{eqnarray*}
U_{C}=d \left( E_{c} -4 \pi P_{a}\right).
\end{eqnarray*}
Here $d$ is the size of the gap in a $LC$ circuit,  $E_{c}$ is the electric field induced by charges at the edges of a $LC$ circuit, and $P_a$ is the polarization of the medium inside the capacitor.   The electric field in the capacitor is
\begin{eqnarray*}
E_{c}=2ES_c=4\pi \sigma=4\pi \left( \frac{q}{S_{C}}\right),
\end{eqnarray*}
where $S_c$ is the area of the metallic plates formed at the end of the $LC$ circuit.

Taking into account the expression for $E$, we see that the expression for $U_{C}$  has the following form:
\begin{eqnarray*}
U_{C}=d 4 \pi \left( \sigma - P_{a}\right)=4 \pi d \left( \frac{q}{S_C} - P_{a}\right).
\end{eqnarray*}

The potential difference on the inductance and the electromotive force can be expressed as:
\begin{eqnarray*}
U_{L}=\frac{L}{c^2}\frac{d I}{d t},~~~~~  \mathcal{E}_{in}=-\frac{1}{c}\frac{d \Phi}{d t},
\end{eqnarray*}
which means that the equation describing this $LC$-circuit reads
\begin{eqnarray*}
4 \pi d \left( \frac{q}{S_C} -P_a \right) =-\frac{L}{c^2}\frac{d I}{d t}- \frac{1}{c}\frac{d \Phi}{d t},
\end{eqnarray*}
that is,
\begin{eqnarray*}
\frac{d I}{d t}+ \frac{4 \pi d c^2}{L} \left( \frac{q}{S_c} -P_a \right)= -\frac{c}{L}\frac{d \Phi}{d t}
\end{eqnarray*}
Recall that the capacity is given by:  $C=S_{C}/(4 \pi d)$, which allows us to make the following substitution:
\begin{eqnarray*}
\frac{d I}{d t}+ \frac{c^2}{C L} \left( q-S_{C} P_{a} \right) = -\frac{c}{L}\frac{d \Phi}{d t}
\end{eqnarray*}
From the expression for Thompson's frequency,  $\omega_{T}^{2} = (LC/c^2)^{-1}$, we obtain system of equations
\begin{eqnarray*}
\frac{d I}{d t}+ \omega_{T}^{2} \left( q-S_{C} P_{a} \right) &=& -\frac{c}{L}\frac{d \Phi}{d t},\\
\frac{d q}{d t}&=&I.
\end{eqnarray*}
If we consider a system with loss we add the potential drop due to resistance, $U_{R}=I R$, to the system and the governing equations will read:
\begin{eqnarray}
\frac{d I}{d t}+ \frac{R}{L} I+ \omega_{T}^{2} \left( q-S_{C} P_{a} \right) &=& -\frac{\pi a^{2}c}{L}\frac{d H}{d t}\nonumber \\
\frac{d q}{d t}&=&I. \label{Gap:Material}
\end{eqnarray}
We know that the polarization effects due to the presence of active atomic inclusions has the form:
\begin{eqnarray*}
P_{a}=  n_{a}( \rho^*_{12} d_{12}+ \rho_{12} d_{21}),
\end{eqnarray*}
where $n_a$ is the density of the atomic population, $\rho_{21}$ is an off-diagonal element of the density matrix and $d_{12}$ is the dipole moment.  Since these atoms are contained within the capacitor, we can express the atomic density as:
\begin{eqnarray*}
n_{a}=\left ( \frac{N_{a}}{S_{C} d }\right ),
\end{eqnarray*}
where $N_a$ is the population and $S_c d$ is the volume of the capacitor.  Thus, the polarization term that comes from atomic inclusions placed inside the  capacitor of the $LC$ circuit can be expressed as:
\begin{eqnarray*}
S_{C} P_{a}=\frac{N_a}{d}( \rho^*_{12} d_{12}+ \rho_{12} d_{21}).
\end{eqnarray*}
Since the size of the "capacitor" is small it follows that the polarization contribution is quite large.
Recalling that the magnetization, $M$, can be expressed as a function of the current in the $LC$-circuit,
$M=\varkappa I (t).$
The material equation determined by these nanostructures with inclusions placed inside the capacitor is therefore:

\begin{eqnarray}
\frac{\partial^2 M_{ns}}{\partial t^2} +\gamma \frac{\partial M_{ns}}{\partial t} +\omega^2_T M = -b \frac{\partial^2 H}{\partial t^2} +\omega^2 _T \alpha \frac{\partial P_a}{\partial t }, \label{Magnetization:comp}
\end{eqnarray}
where $b=\pi a^2 c/L \kappa$, $\alpha = \kappa/d$, and $\gamma =R/L\kappa$.

Now, as in the previous case we will be considering structures responding as an $LC$-circuit to the magnetic field and these structures can have a resonant response to external electric field.  The polarization effects  comes from plasmonic oscillations and the geometry of the nanostructures:
\begin{eqnarray}
\frac{\partial^2 P_{ns}}{\partial t^2} + \delta \frac{\partial P_{ns}}{\partial t} + \omega^2_D P_{ns} = \frac{\omega^2_p}{4\pi} E,
\label{Polarization:nano}
\end{eqnarray}
where $E$ is the external electric field, $P_{ns}$ is the polarization contribution from the nanostructures, $\delta$ is the losses due to plasmonic oscillations, $\omega_D$ is the frequency of the dimensional quantization that comes from the geometry of the resonator, and $\omega_p$ is the plasma frequency.

Now that we have the material equations for this case we need to derive the slowly-varying envelope equations.
From Eq.~(\ref{Magnetization:comp}), we take the Fourier transform to see that:
\begin{eqnarray}
\hat{M} =\frac{\beta \omega^2}{\omega^2_T -\omega^2 - i\omega \gamma} \hat{H} -\frac{i\omega \omega^2_T \alpha }{\omega^2_T -\omega^2 -i\omega \gamma} \hat{P}_a
\end{eqnarray}
Thus,
\begin{eqnarray}
\hat{B} = \left( 1+ \frac{4\pi \beta \omega^2}{\omega^2_T -\omega^2 - i\omega \gamma}  \right)  \hat{H} -\frac{i4 \pi \omega \omega^2_T \alpha}{\omega^2_T -\omega^2 -i\omega \gamma} \hat{P}_a,
\end{eqnarray}
and we see that the effective permeability of the system, $\tilde{\mu}(\omega)$, is the same as in the case of mean-field amplification.  By taking the Fourier trasform of Eq.~(\ref{Polarization:nano}) we obtain
\begin{eqnarray}
\hat{D} = \left( \epsilon_0(\omega) + \frac{\omega^2_p}{\omega^2_D -\omega^2 -i\delta \omega}\right) \hat{E}
\end{eqnarray}
to see that the effective permittivity, $\tilde{\varepsilon}(\omega)$, is also the same as that for mean-field amplification, which is determined by the nano-structures.
Recalling the Fourier transform of Maxwell's equations we have:
\begin{eqnarray*}
k^2 \hat{E} -\frac{\omega^2 \epsilon_0(\omega)}{c^2} \hat{E} = \frac{4\pi \omega^2 }{c^2} \hat{P}_{ns} +\frac{4\pi \omega k}{c}\hat{M} \\
k^2 \hat{H} -\frac{\omega^2 \mu_0(\omega)}{c^2}\hat{H} =\frac{4\pi \omega^2 \varepsilon_0(\omega)}{c^2}\hat{M} +\frac{4\pi \omega k}{c}\hat{P}_{ns}
\end{eqnarray*}
where $\hat{M}$ comes from the nanostructures.  Note that the polarization effects from the active atoms interact mainly with the local electric field inside the capacitor, which is much stronger than the external electric field.    Thus, we have
\begin{eqnarray}
\left[k^2  -\frac{\omega^2\tilde{\varepsilon}(\omega) \tilde{\mu}(\omega)}{c^2}\right]\hat{E}&=& -\frac{i 4\pi \omega^2 k \omega^2_T \alpha}{c(\omega^2_T -\omega^2 -i\omega \gamma)}\hat{P}_a \\
\left[k^2 -\frac{\omega^2 \tilde{\varepsilon}(\omega)\tilde{\mu}(\omega)}{c^2}\right] \hat{H} &=& -\frac{i (4\pi)^2 \omega^3 \epsilon_0(\omega) \omega^2_T \alpha}{c^2(\omega^2_T -\omega^2 -i\omega\gamma)} \hat{P}_a
\end{eqnarray}
Thus, the expansion of the bracketed terms, and the eventual resulting slowly-varying envelope approximation is similar to the previous case.
\begin{eqnarray}
i\left[\frac{\partial}{\partial x} +\frac{1}{v_g}\frac{\partial}{\partial t} +\Gamma  \right] \mathcal{E} = - \frac{i 4\pi \omega^2_0 k_0 \omega^2_T \alpha}{c(\omega^2_T -\omega^2_0 -i\omega_0 \gamma)}\mathcal{P}_a \\
i\left[\frac{\partial}{\partial x} +\frac{\Phi}{v_g}\frac{\partial}{\partial t} +\Gamma -i\Omega \right] \mathcal{H} =-\frac{i (4\pi)^2 \omega^3_0 \epsilon_0(\omega_0) \omega^2_T \alpha}{c^2(\omega^2_T -\omega^2_0 -i\omega_0\gamma)} \mathcal{P}_a
\end{eqnarray}
In addition to the above material equations we must recall that the time derivative of the local electric field is directly proportional to the magnetization that results from the nanostructures, that is,
\begin{eqnarray}
\frac{\partial E_{loc}}{\partial t} = \nu M_{ns},~~~~~~\nu=\frac{4\pi}{S_c \kappa}
\end{eqnarray}
Thus, we have that the Fourier transform of the local electric field can be written as the direct sum of the Fourier transforms of the external electric field and polarization contribution from the active atomic inclusions, that is,
\begin{eqnarray}
\hat{E}_{loc} = \frac{i\nu \beta \omega}{\omega^2_T -\omega^2 -i\omega \gamma} \hat{H} +\frac{ 4\pi  \omega^2_T \alpha}{\omega^2_T -\omega^2 -i\omega \gamma} \hat{P}_a
\end{eqnarray}
where as before $H$ is the external magnetic field.

We now add to these slowly-varying equations those for the polarization and relative atomic population as determined in the previous case.  The difference we must keep in mind is that these amplifying structures are embedded into the nano-structures, and therefore the electric field in these equations is the local electric field rather than the external. The
slowly-varying governing equations for the amplifying material:
\begin{eqnarray*}
\frac{\partial \mathcal{P}_a}{\partial t} = \frac{2 i}{\hbar}N \mathcal{E}_{loc} - i(\omega_{12} -\omega_0) \mathcal{P}_a \\
\frac{\partial N}{\partial t} =-\frac{i}{\hbar} (\mathcal{E}^*_{loc} \mathcal{P}_a + \mathcal{E}_{loc} \mathcal{P}^*_a),
\end{eqnarray*}
where as before $\omega_{12}$ corresponds to the transition frequency and $\omega_0$ is carrier frequency.  Thus, $\omega_{12}-\omega_0$ is the detuning.  To simplify this expression let us replace $\mathcal{E}_{loc} \rightarrow -i\mathcal{E}_{loc}$.

Thus the slowly-varying envelope approximation for the governing equations are:
\begin{eqnarray}
i\left[\frac{\partial}{\partial x} +\frac{1}{v_g}\frac{\partial}{\partial t} +\Gamma  \right] \mathcal{H} =-\frac{i (4\pi)^2 \omega^3_0 \epsilon_0(\omega_0) \omega^2_T \alpha}{c^2(\omega^2_T -\omega^2_0 -i\omega_0\gamma)} \mathcal{P}_a \\
\frac{\partial \mathcal{P}_a}{\partial t} = \frac{2}{\hbar}N (\eta \mathcal{H} + \lambda \mathcal{P}_a) - i(\omega_{12} -\omega_0) \mathcal{P}_a \\
\frac{\partial N}{\partial t} =-\frac{1}{\hbar} ((\eta \mathcal{H} + \lambda \mathcal{P}_a)^* \mathcal{P}_a + (\eta \mathcal{H} + \lambda \mathcal{P}_a) \mathcal{P}^*_a),
\end{eqnarray}

that is
\begin{eqnarray}
\left[\frac{\partial}{\partial x} +\frac{1}{v_g}\frac{\partial}{\partial t} +\Gamma \right] \mathcal{H} =\tilde{a} \mathcal{P}_a \\
\frac{\partial \mathcal{P}_a}{\partial t} = \frac{2}{\hbar}(\eta N  \mathcal{H} + \lambda N \mathcal{P}_a) - i(\omega_{12} -\omega_0) \mathcal{P}_a \\
\frac{\partial N}{\partial t} =-\frac{1 } {\hbar} (\eta^* \mathcal{H}^* \mathcal{P}_a + \eta \mathcal{H} \mathcal{P}^*_a +(\lambda^*+\lambda)|\mathcal{P}_a|^2),
\end{eqnarray}
where $\tilde{a}$ is $i$ times the complex coefficient in front of the polarization from the previous equations.  This is done to make notation simpler.

To determine the governing equations in dimensionless variables let us first multiply the first equation in this system of equations by $\eta$ and let us relabel $\eta H=H$ for convenience.  Thus we will have:
\begin{eqnarray}
\left[\frac{\partial}{\partial x} +\frac{1}{v_g}\frac{\partial}{\partial t} +\Gamma \right] \mathcal{H} =a \mathcal{P}_a \\
\frac{\partial \mathcal{P}_a}{\partial t} = \frac{2}{\hbar}( N  \mathcal{H} + \lambda N \mathcal{P}_a) - i(\omega_{12} -\omega_0) \mathcal{P}_a \\
\frac{\partial N}{\partial t} =-\frac{1 } {\hbar} ( \mathcal{H}^* \mathcal{P}_a +  \mathcal{H} \mathcal{P}^*_a +(\lambda^*+\lambda)|\mathcal{P}_a|^2),
\end{eqnarray}
where $a=\eta \tilde{a}$.  Now, let us write $a=a' + i a''$ and introduce the following dimensionless
variables:  $H=H_0 h$, $P_a = P_0 \rho$, $N=N_0 n$ , $x= z/z_0$, $\tau = t/\tau_0$.  Thus in dimensionless
variables the governing equations read
\begin{eqnarray}
h_z + h_\tau + \gamma h = (1+i\tilde{\alpha}) \rho \\
\rho_\tau = n h +\tilde{\lambda} n \rho + i\Delta \omega \rho \\
n_\tau = -\frac{1}{2} ( h^* \rho +h \rho^* + (\tilde{\lambda}^* +\tilde{\lambda}) |\rho|^2)
\end{eqnarray}
where $z_0 = \tau_0 v_g$, $P_0=N_0$, $\gamma= \Gamma z_0$, $\tilde{\lambda} =\lambda/(\tau_0 v_g a')$, $\tilde{\alpha} =a''/a'$, $\Delta \omega = \tau_0 (\omega_{0} -\omega_{12})$. This is a new system of equations describing electromagnetic field interaction with nanostructures and the interaction of nanostructures with gain material via the induced local fields. Note that the usual conservation law holds in this case:  $|\rho|^2 +n^2 =1$.

\subsection{Front velocity computation}

To compute velocity of a steady wave solution we will use technique described in  Near the excited state $n\approx 1$.  This reduces the system of equations to
\begin{eqnarray}
h_x +h_t + \gamma h = (1+i \tilde{\alpha}) \rho \\
\rho_\tau = h +\tilde{\lambda} \rho -i \Delta \omega \rho
\end{eqnarray}

In a co-moving frame $\tau=t-x/v$, $z=x$

\begin{eqnarray}
h_z -\beta h_\tau +\gamma h = (1+i\tilde{\alpha})\rho \\
\rho_\tau =h +\lambda \rho -i\Delta \omega \rho
\end{eqnarray}

where $\beta =(1-v)/v$.

The dispersion relation for this equation therefore reads:
\begin{eqnarray}
k(\omega) =-\beta \omega +i\gamma +\frac{1+i\tilde{\alpha}}{\omega -\Delta \omega +i \tilde{\lambda}}
\end{eqnarray}
We know from the previous case that to obtain the velocity of the front we must find the zeros of the first derivative of the dispersion, which in this case are:
\begin{eqnarray}
\omega_{\pm} =\Delta \omega +i\tilde{\lambda} \pm i\sqrt{\frac{1+i\tilde{\alpha}}{\beta}}
\end{eqnarray}
Using these to evaluate the imaginary part of the phase as before we obtain:
\begin{eqnarray}
\mathrm{Im}[k(\omega_{\pm})] =-\beta \lambda' +\gamma \mp 2 \mathrm{Re}[i\sqrt{1+i\tilde{\alpha}}] \sqrt{\beta}
\end{eqnarray}
where $\tilde{\lambda}= \lambda' + i\lambda ''$.
Thus, the velocity is given
\begin{eqnarray}
\sqrt{\beta} =\frac{-2 \mathrm{Re}[i\sqrt{1+i\tilde{\alpha}}]+\sqrt{4(\mathrm{Re}[i\sqrt{1+i\tilde{\alpha}}])^2 +4 \lambda' \gamma}}{2 \lambda '}
\end{eqnarray}
If we expand around $\lambda'=0$ we see that up to second order
\begin{eqnarray}
\sqrt{\beta}\approx \frac{\gamma}{2\mathrm{Re}[i\sqrt{1+i\tilde{\alpha}}]} -\frac{\gamma^2 \lambda'}{8 \mathrm{Re}[i\sqrt{1+i\tilde{\alpha}}]^3} +\frac{\gamma^3 (\lambda')^2}{16 \mathrm{Re}[i\sqrt{1+i\tilde{\alpha}}]^5}
\end{eqnarray}
From this we see that when $\lambda'=0$ the velocity is the same as the most basic case if $\tilde{\alpha}=0$ as well, that is $\beta = \gamma^2/4$.

\section{Conclusion}

In this paper we theoretically studied the propagation of a steady state ultrashort pulse in a dissipative medium. We considered the case when  the medium  consists of lossy  metallic nanostructures embedded to a  gain material and the case when the gain material is embedded directly to a nanostructures. We derived a system of governing equations and  found analytic expressions for the  velocity and spatial period of an optical  pulse coupled with the polarization wave. We also proposed a simple technique that describes the shape of this pulse using a simple system of ordinary differential equations.

\section*{Acknowledgment}

We would like to thank  Andrei Sarychev and Alexander Kildishev  for
enlightening discussions.  A.I.M appreciates  support and
hospitality of the University of Arizona Department of Mathematics
during his work on this manuscript. This work was partially
supported by NSF (grant DMS-0509589), ARO-MURI award 50342-PH-MUR
and State of Arizona (Proposition 301), RFBR (grant No.
09-02-00701-a).

\end{document}